\documentclass[preprint,proceedings]{rmaa}

\usepackage{paralist}

\usepackage{psfrag,color}

\SetYear{2005} \SetConfTitle{LARIM}

\title{METALLICITY GRADIENT AND THE HYBRID FORMATION SCENARIO FOR
EARLY-TYPE GALAXIES}
\author{
  Ricardo L.C. Ogando,\altaffilmark{1,2}
  Marcio A.G. Maia,\altaffilmark{2}
  Paulo S. Pellegrini,\altaffilmark{2}
  Cristina Chiappini,\altaffilmark{3}
  Ricardo P. Schiavon,\altaffilmark{4}
  and L.N. da Costa\altaffilmark{2}}
\altaffiltext{1}{Instituto de F\'\i sica, Universidade Federal do
Rio de Janeiro, Rio de Janeiro, Brazil.(ogando@if.ufrj.br)}
\altaffiltext{2}{Observat\'orio Nacional, MCT, Rio de Janeiro,
Brazil.} \altaffiltext{3}{Osservatorio Astronomico di Trieste,
INAF, Trieste, Italy.} \altaffiltext{4}{Department of Astronomy,
UVa, Charlottesville, USA.}

\shortauthor{Ogando, et al.}
\shorttitle{Metallicity Gradient And The Hybrid Formation
Scenario}

\fulladdresses{

\item C. Chiappini: Osservatorio Astronomico di
Trieste, INAF, Via G. B. Tiepolo 11, 34124 Trieste, Italy
(\email{chiappini@ts.astro.it}).

\item L.N. da Costa, M.A.G. Maia and P.S. Pellegrini:
Observat\'orio Nacional, MCT, Rua Jos\'e Cristino, 77, Rio de
Janeiro, 20921-400, RJ, Brazil   (\email{ogando, maia, pssp,
ldacosta@on.br}). 

\item R. L. C. Ogando: Instituto de F\'\i sica, Universidade
Federal do Rio de Janeiro, Centro de Tecnologia, Sala A-329,
Cep:21941-972, Brazil  (\email{ogando@if.ufrj.br}).

\item R. P. Schiavon: Department of Astronomy, University of
Virginia, P.O. Box 3818, Charlottesville, VA
(\email{ripisc@virginia.edu}).
 }

\listofauthors{  R. L. C. Ogando, M. A. G. Maia, C. Chiappini, P.
S. S. Pellegrini, R. P. Schiavon, \& L. N. da Costa}
\indexauthor{Ogando, R.L.C.} \indexauthor{Maia, M.A.G.}
\indexauthor{Chiappini, C.} \indexauthor{Pellegrini, P.S.}
\indexauthor{Schiavon, R.P.} \indexauthor{da Costa, L.N.}

\abstract{We present radial gradients of the Lick index $Mg_2$ for
40 early-type galaxies. In plots of $\triangle Mg_2$ versus mass
indicators, such as $\log\sigma$, the lower boundary of the points
distribution may be populated by galaxies which predominantly
formed by monolithic collapse. Galaxies showing flatter gradients
at higher masses could represent objects which suffered important
merging episodes. Thus, our results support a hybrid formation
scenario. To remove possible age effects, we computed metallicity
gradients ($\triangle [Z/H]$) using $Mg_2$ and $H\beta$ indices
for an $[\alpha/Fe]=0.3$ single stellar population model. The
conclusions remain the same.}

\resumen{We present radial gradients of the Lick index $Mg_2$ for
40 early-type galaxies. In plots of $\triangle Mg_2$ versus mass
indicators, such as $\log\sigma$, the lower boundary of the points
distribution may be populated by galaxies which predominantly
formed by monolithic collapse. Galaxies showing flatter gradients
at higher masses could represent objects which suffered important
merging episodes. Thus, our results support a hybrid formation
scenario. To remove possible age effects, we computed metallicity
gradients ($\triangle [Z/H]$) using $Mg_2$ and $H\beta$ indices
for an $[\alpha/Fe]=0.3$ single stellar population model. The
conclusions remain the same.}
\addkeyword{Galaxies:\ abundances} \addkeyword{Galaxies:\
elliptical and lenticular} 
\begin{document}
\maketitle

\section{Introduction}

Two distinct galaxy formation scenarios have been proposed to
explain the build-up of elliptical galaxies: Monolithic
Dissipative Collapse (MDC) and Hierarchical Clustering (HC). In
the MDC model a galaxy forms rapidly from a primordial cloud. An
important prediction of this scenario is that early-type galaxies
should present a metallicity gradient (MG) and that this MG should
correlate with the object mass. In the HC scenario a galaxy is
formed by coalescence of smaller objects. No MG is expected, since
the successive mergers along the history of a galaxy are expected
to remove this feature. Results of numerical simulations
\citep{Kob04} show that, even in a HC scenario, both mechanisms
are necessary to explain the observed behavior of MG.
\citet{Oga05} determined Mg$_2$ radial gradients ($\triangle
Mg_2$) for a sample of early-type galaxies and their major
conclusion points to the necessity of a hybrid scenario to explain
the observations. Here we repeat the analysis including MG for 10
more objects.

\section{Sample and Observations}

Galaxies were selected from the ENEAR survey \citep{daC00}, which
contains a database of photometric \citep{Alo03} and spectroscopic
\citep{Weg03} parameters for a sample of early-type galaxies,
which are representative of the nearby Universe. The spectroscopic
data were obtained with the 1.52 m telescope at ESO. The 1-D
extractions were made as follows: the central aperture has 3
pixels ($\simeq$2.5"). Successive lateral apertures are set in
such a way that their central pixels are the outermost pixels of
the previous apertures, keeping the same size. This process
continues while a $S/N \geq20$ is obtained for the spectra.
Conversion to Lick system was made by: degrading spectral
resolution and corrections for offsets and velocity dispersions. A
linear fit weighted by the indices' errors was used to measure its
variation as a function of $\log(r/r_e^*)$, where $r_e^*$\ is
$r_e$\ corrected for the galaxy ellipticity
($r_e^*=r_e(1-\epsilon)^{-1/2}$). For the analysis in the next
section, we also included the data obtained by \citet[]{Car93,
Car94a, Car94b}. The set of galaxies in the present analysis has
similar $\sigma$ distribution to that of ENEAR to a 97\% of
confidence level given by the Kolmogorov-Smirnov test.

\section{Results and Discussion}

To test one of the predictions of MDC model, the dependence of MG
with galaxy mass we assume that $Mg_2$ line index predominantly
reflects the metallicity. In Fig.1 (panel a) we plot $\triangle
Mg_2/ \triangle \log\ r/r_e^*$ versus $\log\ \sigma$. This plot
reveals that at least part of the galaxies located at the lower
boundaries of the point distribution show an $Mg_2$ gradient which
increases with $\sigma$ presenting an $MG-\sigma$ relation similar
to that in the simulations of \citet{Kob04}, indicating the
dominance of monolithic collapse. The galaxies which occupy the
locus of low MGs, according to the same simulations coincide with
products of mergers.
\begin{figure}[!t]
  \includegraphics[angle=-90, width=\columnwidth]{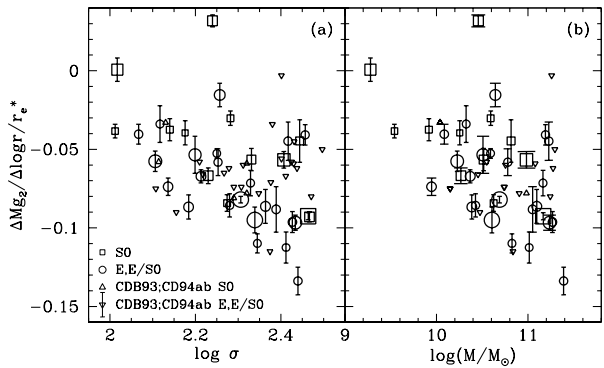}
  \caption{Radial $Mg_2$ gradients vs.\@ central velocity dispersion
  $\sigma$ (panel a) and versus $Mass$ (panel b). We include the data
  by \citet[]{Car93, Car94a, Car94b}. Their $E$s and $E/S0$s
  are indicated by ``$\nabla$'', the $S0$s by ``$\triangle$''.
  The size of the symbol is proportional to the fraction of $r_e$ in
  which we measured the $\triangle Mg_2$ (the bigger ones means
  $r/r_e\approx 1$).}
  \label{fig:fig1}
\end{figure}
In Fig.1 (panel b) we plot the $\triangle Mg_2$ vs.\@ $\log\ Mass$
($Mass \propto r_e \sigma^2$), where $r_e$ is the effective
radius. This plot is very similar to that of panel (a) and
reinforces the hypothesis of existence of objects reflecting the
dominance of collapse or merger in the process of galaxy
formation.

To remove possible age effects, we estimate [Z/H] gradients using
$Mg_2$ and $H\beta$ indices and single stellar population models
of \citet{Tho03} for [$\alpha$/Fe]=+0.3. [Z/H] are estimated for
$r_e/8$ and $r_e/2$ from the $Mg_2$ and $H\beta$ values
interpolated in the linear index vs.\@ $\log r/r_e$ relations. In
Fig.2 we display the $\triangle [Z/H]/ \triangle \log\ r/r_e^*$
versus $\sigma$ (panel a) and $\triangle [Z/H]/ \triangle \log\
r/r_e^*$ versus $Mass$ (panel b) for 25 galaxies. The overall
distribution of the data points is very similar to that in Fig.1.
\begin{figure}[!t]
  \includegraphics[angle=-90, width=\columnwidth]{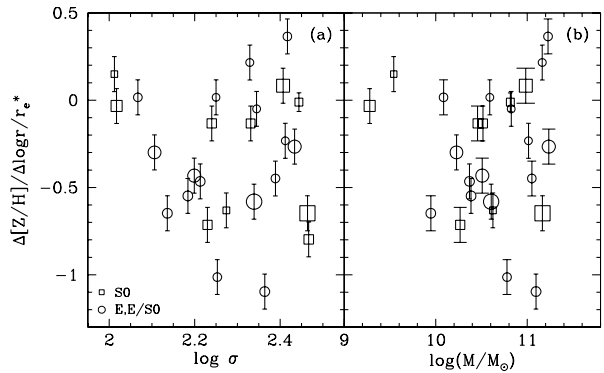}
  \caption{Radial $\triangle [Z/H]$ as a function of galaxy central velocity
   dispersion $\sigma$ (panel a) and $Mass$ (panel b). The number of points is
   smaller in this plot because some line indices measurements of $Mg_2$ and
   $H\beta$ used to determine $[Z/H]$ fall outside the grid of single stellar
   population models by \citet{Tho03}.}
  \label{fig:fig2}
\end{figure}
These results give observational support to the hybrid scenario
and show that we can interpret galaxy formation not as an
exclusive dominance of either MDC or HC scenarios. Each galaxy has
its particular formation history depending on the merger events of
its building blocks, their nature (predominantly gas or stars) and
their efficiency to collapse. The location of a galaxy in a
$MG-\sigma$ diagram may be a useful tool to infer the relative
importance of mergers or collapse to its formation.
\acknowledgements R.L.C.O.\@ acknowledges CNPq Fellowship;
M.A.G.M.\@ CNPq grants 301366/86-1, 471022/03-9 and 305529/03-0,
P.S.P.\@ CNPq grant 301373/86-8.

\end{document}